# Ultrathin, flexible and MRI-compatible microelectrode array for chronic single units recording within subcortical layers

Scott Greenhorn[1], Véronique Coizet[2], Victor Dupuit [1], Bruno Fernandez [1], Guillaume Bres[1], Arnaud Claudel[1], Pierre Gasner[1], Jan M. Warnking[2], Emmanuel L. Barbier[2], Cécile Delacour[1]*

1. *Institut Néel, University Grenoble Alpes, CNRS, Grenoble INP, 38000 Grenoble, France*

2. Univ. Grenoble Alpes, Inserm, U1216, CHU Grenoble Alpes, Grenoble Institut Neuroscience, 38000, Grenoble, France

*Corresponding author: cecile.delacour@neel.cnrs.fr

**Abstract.** Current techniques of neuroimaging, including electrical devices, are either of low spatiotemporal resolution or invasive, impeding multiscale monitoring of brain activity at both single cell and network levels. Overcoming this issue is of great importance to assess brain's computational ability and for neurorehabilitation projects that require real-time monitoring of neurons and concomitant networks activities. Currently, that information could be extracted from functional MRI when combined with mathematical models. Novel methods enabling quantitative and long-lasting recording at both single cell and network levels will allow to correlate the MRI data with intracortical activity at single cell level, and to refine those models. Here, we report the fabrication and validation of ultra-thin, optically transparent and flexible intracortical microelectrode arrays for combining extracellular multi-unit and fMRI recordings. The sensing devices are compatible with large-scale manufacturing, and demonstrate both fMRI transparency at 4.7 T, and high electrical performance, and thus appears as a promising candidate for simultaneous multiscale neurodynamic measurements.

**Teaser.** Traditional neuroimaging methods struggle with low resolution or invasiveness, hindering our grasp of brain activity. Long lasting and MRI compatible intracortical microelectrode arrays could bring a significant breakthrough to investigate deep brain activity. The reported conformable and f-MRI compatible devices is meant to that goal, enabling single spike and multichannel recordings within deep brain networks. We believe these devices open the road for multiscale neurodynamic measurements, by combining electrophysiology and functional MRI.

# MAIN TEXT

## Introduction

A number of different platforms exist to picture neuronal activity at multiple temporal and spatial scales but none gathers all in once which is a common shortcoming to follow the long range and fast dynamics of neural communications for both fundamental and clinical applications. Indeed, neuronal signals are widely spread both in frequency and spatial domains, ranging from microscale fast unitary spikes to slow oscillatory waves at the brain level, while the available neurorecording methods are either limited by their spatial and/or temporal resolution (e.g. MRI, EEG, PET) or by their invasiveness (e.g. MEA, fluorescent imaging) impeding whole brain imaging with single cell resolution.[1]

Together MRI and microelectrode array leverages the strengths of both of these two functional neuroimaging methods, providing the required wide field of view and high spatiotemporal resolution for monitoring (or stimulating) individual neurons in real-time with large network activation. Such achievement would allow to track long range neural communications and provide valuable insights on brain physiology and function and for translational research. Thus, finding MRI-compatible electrodes has been an old scientific and technological subject of research.

Pioneering work successfully demonstrated real-time acquisition of local field potential or unitary spike during functional MRI sequences (blood oxygenation level-dependent, epi or flash MRI sequences)[2–4] and several achievements were reported since. To give a few examples, simultaneous fMRI acquisition and electrode recordings enabled mapping functional connectivity and adaptability of the brain,[5] providing multimodal visualization of neural changes associated with learning and task execution.[6] Also, dual electrophysiological recordings helped to better understand the relationship between the fMRI blood oxygenation level-dependent (BOLD) signal, which is widely used as a surrogate for neural activity, and the underlying electrical signals.[2,3,5,7] However, current electrode designs generate MRI artefacts, preventing an MRI observation at sites of interest, and raise MRI safety issues. To enable MRI observations in implanted patients, electrodes inducing little to no artefacts in MRI imaging have been proposed.[4,8,9] These electrodes enable MRI follow-up of the acute and chronic states of the tissue located near the electrode, after its surgical implantation.

Several microelectrode designs and materials have been used such as glass-coated platinum-iridium electrodes,[2] metallic (tungsten[10], platinum–iridium[11] or platinum–tungsten[12]) wires, graphene-encapsulated microwires[9], and silicon multichannel microelectrodes.[13,14] Most of these electrodes were good candidates, showing artefact-free properties up to MRI field strengths of 21 T without statistical difference in the quality of the electrophysiological recordings, but were limited in the number of recording channels or the probe rigidity eventually leading to poor biocompatibility and long-term recording issues. Mechanical mismatch with cells and tissues elicits a sustained inflammation and immune response around the electrode over time, leading to neuronal degradation and glial scar formation and loss of electrophysiological signals.[15] In addition to possible corrosion of electrode materials, these issues cause degradation of device performance over time during chronic implantation. Therefore, the implementation of microelectrode arrays on flexible and ultra-thin materials has appeared as a promising approach for sustainable neural implants.[16] The low quantity of inserted materials, its conformability and its bendability ensure a soft and close contact with tissues and cells to satisfy biocompatibility issues, while offering multichannel recording and charge injection possibilities during wide-field brain imaging inside MRI systems.

Flexible microelectrode arrays have been developed using carbon,[4,8,9] organic polymer,[17–19] liquid and solid metal[19–23], porous Pt,[24] graphene[25–28] and MXene[29,30] as the sensitive or coating electrode material. Among them gold electrodes on Parylene have been shown to record unitary spike from individual sensory and LG neurons,[31,32] EEG signals as epidermal electrodes,[33] and have been able to stimulate neural activity[7,34] while allowing low-artefact MRI imaging.[35] Carbon nanotube fibers have been used before to achieve a reduced electrode footprint in MRI.[5,9,15] However, these devices were less flexible and had lower longevity than comparable polymer probes, and required an additional carrier to aid the insertion process.

Here, the advances in flexible electronics are leveraged for high throughput processes of ultrathin and conformable microelectrode arrays, enabling multichannel recording of single units from individual subcortical neurons in rodents. Both transparent gold and graphene microelectrodes can be implemented that enable optical imaging and artefact-free MRI acquisitions (shown here at 4.7T). The interactions between MRI RF and gradient magnetic fields and the MEA are assessed in detail, in terms of charge injection with the microelectrode and signal to noise ratio during electrical MEA recordings, in terms of MR image artefacts, and in terms of RF-induced heating in the presence of the implant. Beyond the new possibilities for correlative fMRI and electrophysiological recordings, the ability to use the microelectrode not for recording only but also for local stimulation in combination with fMRI, is of great interests to study epilepsy or to alleviate movement and behavioral disorders on implanted subject with deep brain stimulation (DBS).[4,11] But this raises important security issues (e.g. heating at the electrode tips, induced currents, magnetic field–induced implant movement) which prevent the use of fMRI in many cases.[36] Here, the proposed Flex-MEA gathers several advantages such as micrometer size, a low susceptibility and a negligible amount of metallic materials that help reducing MR-safety related risks significantly, and could ensure safe MRI in implanted patients and open a way for expanding this tool in future research and clinical cares. Also, the reported manufacturing process is compatible with CMOS microelectronic platform for large scale implementation at manufacturing levels to be readily generalized beyond academic laboratories.

**Results**

1. **Microfabrication of the MEA**

Arrays of ultrathin, flexible, conformable and optically transparent microelectrode arrays are fabricated with standard CMOS microfabrication platforms, and using gold and graphene as the microelectrode material (**figure 1**). Figure 1a describes the fabrication steps (details in materials and methods). Briefly, the substrate is coated with a first layer of Parylene-C (20 µm). The metallic wires and microelectrodes are obtained by the lift-off in acetone of a 5 nm – 45 nm thick titanium-gold layer evaporated in high vacuum over a photoresist previously patterned by laser lithography. A second layer of Parylene-C (15 µm) is deposited to isolate the metallic lines for liquid operation, and it is etched at recording sites to expose the microelectrode and to outline the 7 mm long device shank. At the end of the process and after all characterizations, the probes are released from the substrate by immersing the wafer in DI water. Optical micrographs assess the overall quality of the devices before and after release (figure 1b). Up to ten separate electrode channels provide redundancy, parallel recording, and allow for the injection of current for intracortical stimulation. For these electrode and lead lengths and impedances, heating effects produced by the interactions of the MRI RF transmit field with the microelectrode array are expected to be negligible.[10] To prevent crosstalk, electrodes are spaced by 200 µm, allowing an area of 1.6 mm² to be mapped.[37] Also, the fabrication process allows the implementation of graphene-based devices which provides additional advantages such as an optically transparency at the recording site to combine electrical and optical imaging. The middle panel figure 1b illustrate the transparency of graphene microelectrode and field effect transistor. Here, the monolayer graphene is preferred to ITO or porous graphene because of its higher transmittance,[38] bendability (REF) and biocompatibility.[39]

Finally, the probes are released from the substrate and assembled with the connectors, the choice of which mostly depends on the final application. Optical micrographs (figure 1c) illustrate several designs of the manufactured neural implants assembled with zero-insertion force (ZIF) connector (Würth Elektronik WR-FPC) which is a standard for chronic recordings.

## 2. Nyquist and Bode plots

The electrochemical impedance and phase spectra of the flexible devices have been investigated for the gold and graphene flexible microelectrode (**figure 2**) and compared with rigid microelectrodes, namely TiN on glass and graphene on sapphire (figures S1 to S3). For each device, the spectroscopy impedance measurements were performed from 1 Hz to 500 kHz with a lock-in preamplifier (MFLI, Zurich Instruments), and using a Pt quasi-reference electrode, voltage-biased at 500 mV in phosphate buffer solution to mimic the cell medium (pH 7.4, resistivity of 56–71 $\Omega$.cm). The Nyquist and Bode plots (blue and green dots) illustrate the frequency-dependence of the impedance modulus and phase of the gold and graphene microelectrodes (figure 2a). The impedance modulus |Z| extracted at 1 kHz is about 44 k$\Omega$ and 67 k$\Omega$ for the flexible graphene and gold microelectrodes. These values are comparable with TiN microelectrodes (46 k$\Omega$) and previous reported values for gold[34] and graphene[25,40–42] microelectrodes, confirming the overall quality of the devices fabricated on flexible polymer.

The data are well described with the equivalent circuit (red line), which includes the serial contributions of the electrolyte spreading resistance, the electrical double layer *ED* and the exposed microelectrode area (figure 2b). Those interfaces are modelled with a resistive element $R_0$ in series with two Randles circuits.[43] The Randles circuits consist of a constant phase element $C_{PE}$ in parallel with a charge transfer resistance ($R_{CT}$), which is in series with a linear diffusion element ($Z_W$). In fact, only the response of graphene microelectrodes fabricated on the rigid sapphire substrate requires to add this Warburg element ($Z_W$) in series with the charge transfer resistance (table S1), suggesting that diffusion dominates the faradic component. This behaviour was also reported on a quartz rigid substrate[40] but not on a flexible substrate for similar size.[41] In all case, the Randles' circuits require a constant phase element $C_{PE} = 1/Q(iw)^n$, revealing non-negligible deviations from an ideal capacitor. This could be due to potential inhomogeneities of the surface reactivity and roughness depending on the *n* exponent value.[44]

At the *EDL* interface, both the faradic and capacitive contributions are found higher on the flexible gold and graphene devices compared to rigid MEA (table 1 and table S1), leading to lower impedance values (figure S1a). The fitting resistances are lower ($R_1$ = 0.7 and 0.98 MΩ versus 97 MΩ respectively) and the capacitance are higher ($C_{PE}$ = 84nS and 98 nS versus 670 pS and 24nS respectively), for the flexible gold and graphene MEA, in comparison with all rigid microelectrodes. Interestingly, the *n* exponent of the constant phase element remains high, ranging between 1 and 0.88 for graphene and gold flex-MEA, underlining the overall homogeneity of the surface roughness and reactivity.[45] The fitting values for the flexible MEA are very close to those reported with conventional process on rigid substrate[25,42], ranging from 0.3 to 84 MΩ and 1.25 µS to 69 nS. This further confirms the high quality of the fabricated flexible microelectrode arrays.

Finally, the flexible microelectrodes showed consistently lower impedances than those of the commercial MEA, using a variety of connection strategies (figure S1b). The direct contact using silver paste was used for both in-vivo electrophysiology and MRI imaging as it provided the most reliable connection and was least destructive to the devices. Its use is however limited to acute recording.

3. **MRI compatibility**

Imaging or recording issues could arise from MRI and MEA acquisitions, due to the interference of implanted electrodes with the static magnetic field and RF field in the MRI system.[46] The characterization of such artefacts is therefore of great interest for both MRI imaging and MEA recording, and to assess the potential of the technology for dual fMRI-MEA imaging. For that purpose, the fabricated devices were immersed in standard phantom gel (1% agarose) used to mimic the brain tissues, and inserted in a 4.7 T Bruker Biospin MRI system (**figure 3**).

First regarding possible MRI issues, data were collected around the implant from a gradient echo sequence, which is sensitive to magnetic susceptibility artefacts (3D flash sequence, detailed in methods). Two configurations have been tested (figure S4). In the first configuration, the implant was placed alone, without any connector or extension, in an orientation perpendicular to the main magnetic field $\vec{B_0}$. In the second configuration, the electrode was placed parallel to $\vec{B_0}$, connected to a flat flexible cable suitable for acute recordings and for biasing the microelectrodes. For each configuration, figures 3a and 3b show the Cartesian orthogonal views through the implant position, to asses any MRI artefacts along the length, the width and the thickness of the probe. The image volume encompasses 128x128x80 voxels (26x26x16 mm³) with an isotropic resolution of 0.2 mm.

For none of the configurations, the inserted probe was visible. Within the two upper images of figure 3a, the interface between the two layers of gel embedding the probe is faintly visible, likely due to the presence of free water. The absence of any visible gradient-echo MRI signal changes in the vicinity of the electrode confirms the absence of any susceptibility artifacts due to the electrode materials. The absence of B1-related artifacts suggests that in the tested configurations there is no RF current induced in the electrode. The bottom left image figure 3a is a section at the position of that interface. No trace of the probe is observed. Only the electrical connection for acute recording is detected proximal to the conformable implant (figure 3b). The location of the connection and extension cable is only given by the signal void corresponding to the presence of the tape and ribbon cable covering the connection. These first observations confirm the MRI compatibility of the manufactured neural probe, which is as expected regarding the size of the device (30μm thick and 200 μm wide), the low quantity of inserted metallic compound, and the low magnetic susceptibility of the substrate.[7] Note that the gold microelectrodes have been used here as they contain the highest quantity of metals compared to graphene which is a single layer of carbon atoms.

Moreover, the perturbation of the magnetic field due to the connected implant has been measured using a $B_0$-mapping sequence (figure S8). Variations of the magnetic field close to the connector are visible but extremely weak (+/- 1.17 μT or +/- 0.25 ppm) and arise from the presence of the electrical connector. No $B_0$ perturbations are detected along the probe near the metallic contact-line or the microelectrodes, being about 5-7 mm away from the connector. Even using gradient-echo echo-planar imaging, a sequence used for fMRI which is extremely sensitive to susceptibility artefacts, the device was undetectable, confirming the possibility to acquire artefact-free fMRI data from tissues at the microelectrode recording sites.

Finally, we have assessed the ability to detect spike-like events during fMRI sequences as well as the impact of injecting current from the microelectrodes on the fMRI acquisition. To that aim, one of the 10 recording channels of the MEA was used to apply voltage pulses (few ms duration and 100 µV amplitude), while the responses of other channels are monitored during a 3D Flash and field map MRI sequences (same condition used previously for MRI imaging). As illustrated within figures 3c and 3d, after removing the MRI artifacts from the recordings in post-processing, the spike-like events are detected during each MRI condition, despite the initially large electrical artefacts caused by the interference from MRI gradients. The bottom trace figure 3c illustrates the MRI-induced signal during a 3D Flash sequence and a reference signal in same condition but without MRI gradient switching (grey and black lines respectively). Electrical artifacts due to MRI gradients remain exactly the same over time, thanks to the consistency of the MRI system, which allows to subtract them in postprocessing and to recover all the spike events even during the strongest artifacts (figure S5). Interestingly, the background noise is approximately the same before and during MRI acquisitions, being about 2nA in each condition. The fMRI sequence does not induce additional noise in the frequency range of interest (300Hz to 10kHz).

Finally, the biasing used for both stimulation and recording have negligible effect on the MRI measurements which is as expected regarding the low quantity of injected charges during brief millisecond period. The difference in the field map is negligible and possibly due to random differences in the automatic field adjustment procedure performed by the machine prior to the acquisition.

The main safety concern for the use of active implants during MRI is the potential for RF-induced heating of tissues in electrical contact with the implant. In order to assess the risk of RF heating, we performed fiber-optic thermometry during MRI using two MRI-compatible high-precision temperature probes (Photon Control Inc., Richmond, BC, USA). The probes have an external diameter of less than 1 mm, a precision of <0.1 K and a sampling rate of 10 measurements per second. Temperature was monitored on an implant with connected wire immersed in Agar Gel, both at the tip of the implant and at the level of the connector, during a T2-weighted Turbo-RARE sequence of a duration of 8 min 24 s exhibiting high SAR (figure 3e). The measurement was repeated after removal of the implant and connecting wire, leaving the temperature sensors in place (figure 3f). The observed temperature rise during the Turbo-RARE sequence with respect to the baseline was fitted with a linear model, allowing the measurement of local specific absorption rate (SAR), assuming the specific heat capacity of the agar gel is close to that of pure water (4184 J/K.kg)[47]. Local background SAR in absence of the implant was 1.90±0.10 W/kg at the implant tip location and 2.97±0.12 W/kg at the connector (measured value ± 95% C.I.). This is comparable to high-SAR sequences in human MRI. In the presence of the implant, the observed SAR was 2.38±0.10 W/kg at the tip and 5.29±0.11 W/kg at the connector. The additional heating observed due to the implant was thus 0.48±0.14 W/kg at the implant tip and 2.32±0.16 W/kg at the connector. The additional heating at the implant tip was about 25% of the background SAR, whereas macroscopic wire-like implants routinely induce additional heating which is many times stronger than the heating due to background SAR[36]. Absolute temperature increases in all conditions remained well below 1 K during the 504 s of acquisition of this of high-SAR MRI sequence, and thus stayed within the accepted safe limits of applications in the brain of less than 2 K[48]. The observed additional heating at the connector was stronger, likely due to the larger wire section in the connecting cable, but the connector not being in contact with tissue in in-vivo applications, this should not have an impact on the safety of the implant.

All together, these characterization assays demonstrate the ability of the flexible MEA to record and stimulate spike-like events during MRI sequence with while keeping all the performance benefits of the both technologies (MEA and MRI). To fully demonstrate the potential of the manufactured implant for correlative fMRI and electrophysiological recordings, we have assessed in which extent the microelectrodes are sensitive enough to detect unitary spikes from individual neurons.

4.  **Single spike recordings in acute condition**

The conformable microelectrodes were tested for the detection of extracellular action potential (namely single-unit or unitary spike) during acute monitoring of intracortical layers (hippocampus and thalamus). A similar method of standard electrophysiological recording with glass pipette and tungsten wire was used for a benchmark study (**figure 4**).

Briefly, rats were anaesthetized with an intra-peritoneal injection of ethyl carbamate (urethane, 1.25 g/kg, Sigma-Aldrich) and placed in a stereotaxic frame, while the body temperature was maintained at 37°C as described in methods. A craniotomy was then performed to allow access to the structures of interest. The electrode was inserted into the brain at the specific Bregma coordinates and lowered until a putative neuron was found. During extracellular voltage excursions, the signal is amplified and band-pass filtered (300 Hz–10 kHz) reducing the background noise down to 20 µV, which is consistent for 30 µm diameter microelectrodes.[49]

Clear signatures of unitary spike are shown within the voltage time-trace figure 4a during resting state (spontaneous activity). The two lower time traces (panel 4a) provide closer views on the recorded spikes which remain highly identical. The millisecond duration, biphasic shape and 20 to 100 ms inter-spike interval (for the left and right traces respectively), are as expected for unitary spikes and bursts from individual neurons. The spike amplitude ranges from 100 to 400 µV peak-to-peak, leading to a high value of the signal to noise ratio S/N=5 – 20. This value outperform the tungsten wire (panel 4b), and is similar to the glass micropipette (panel 4c) the golden standard for electrophysiological recording.[50] Interestingly, the microelectrodes enable to detect weak changes in spikes amplitude during the bursting events which are not observed when using the glass micropipette. This could result in part from the low values of the interfacial resistance and capacitance (figure 2b). Here, the conformability of the microelectrodes improves the electrical contact to the cells compared with rigid microelectrodes.

Evoked activities were detected with the microelectrode arrays (figure S6). For a sound stimulus (hand claps), the microelectrode recorded an oscillatory variation of the local field potential. The signal-to-noise ratio was found to be S/N=2.6 on average, with a baseline noise level of about 20 µV. Also, the system responds with high efficiency to all stimuli, revealing a high detection reliability of local field potential.

5. **Multichannel recordings in chronic condition**

After successful recordings in acute condition, we investigated the efficiency of this technology for chronic implantation and recordings, extending up to 19 days after implantation. Beyond potential detection issues arising in the first week of implantation, mainly due to factors such as the glial scar formation, infection, or inflammatory reactions, this validation study involved more advanced implantation and connection strategies than those employed for acute recording. In this context, the connection between the MEA and the external electronics should ensure reliable and long-lasting electrical contacts, lasting at least several weeks, all while being compatible with freely moving animals and MRI antennas. The optimization of the connector could be explored in a dedicated study which is beyond the scope of the present work. For our current trial, the probe was connected using light and compact zero-insertion force connectors that meet the basic requirements mentioned earlier.

Figures 5 and 6 show the voltage time traces of all the microelectrodes (10 recording channel on the right panels) recorded on days 11 and 19 after implantation, respectively. For each recording period, multichannel recording was performed during a resting period (a), and under auditory (b) and sensory (c) stimuli to capture the spontaneous and evoked activity of the subcortical neurons. For each condition, selected timestamps provide closer views of the detected spike patterns.

The recorded time traces are similar in chronic and acute conditions, displaying typical signatures of both spontaneous and evoked neural responses. The spontaneous activity is characterized by isolated unitary spikes and burst events that are randomly distributed in time. Interestingly, at least two patterns can be distinguished among the microelectrodes on day 11, while the same pattern emerges from the recording on day 19. Few channels (about 1 on day 11) record no activity. This does not result from defective electrode but rather indicates silent neurons. Indeed, the silent channels vary between the recording conditions.

The sound stimuli (hand claps) evoked oscillatory waves of about a few hundred microvolts (100 µV). Furthermore, the neural response is highly reproducible across six sites and for each stimulus. Three electrodes remained silent on day 11. Interestingly, by day 19, all electrodes were activated, but they exhibited lower signal amplitudes (approximately 50 µV). Sensory stimuli (hind paw pinching) also induced a reliable neural response: a first unitary spike followed by a second one of higher amplitude (100-200 µV) delayed by about 100 ms. On day 19, a spontaneous activity pattern followed the sensory-evoked response, which was characterized by isolated spike trains of a few hundred microvolts in amplitude.

Overall, spontaneous bursting patterns may vary based on the microelectrode's location and the excitability of the sensed neurons (figures 4a, 5a, and 6a). Subthalamus neurons located deeper in the SNR (MEA around -7 mm, figure 4) are more likely to express intense burst patterns, whereas activity in upper brain areas within the hippocampus (MEA around -5 mm, figures 5 and 6) might be diminished.

Furthermore, within the same experiment, differences appear between the recording channels (as seen in the right panels of figure 5 and 6), depending on their position along the 1500 µm-long array, as well as the nature and the excitability of the probed neurons. Interestingly, the background noise remains about 20 µV, similar to the value obtained from acute recordings. This value also stays constant over time (both on day 11 and 19), indicating no signs of device degradation. The signal to noise ratio remains comparable to the acute condition (S/N=10 on day 11), with similar values on day 19 (figure 6c). Physiological differences could be the cause of the observed discrepancy between the recording traces.

At a first glance, we do not observe significant differences in term of number of active electrodes, amplitude of the signal recorded and background noise between days 11 and 19. Additional recordings are necessary for a statistical study to assess the long-term performance of the devices and their potential for extended applications (several months), either in clinical or fundamental studies.

### 6. MRI and Tissue footprint after chronic implantation

Further MRI acquisitions were conducted in vivo to evaluate the impact of the implant on MRI images. After seven weeks of implantation, animals were anesthetized with an intraperitoneal injection of ethyl carbamate (urethane, 1.25 g/kg, Sigma-Aldrich), with continuous monitoring of body temperature and cardiac pulsation during MRI recording sessions. The expected position of the probes is indicated by the red cross in figures 7a, 7b, and 7c. While the footprint of three screws securing the cap on the animal's head is clearly visible (right black dot, figure 7a), the trace of the flexible probe (red cross) on the MRI images is weak, consistent with expectations from preliminary acquisitions within the agar gel (figure 3).

However, the signal-to-noise ratio of the MRI images remains low, likely stemming from the multichannel connector used for chronic recordings. Further developments are still needed for an MRI-compatible connector, which was the limiting factor in our case for multichannel recording during chronic applications including awake and freely-moving animals.

At the end of the experiment, astrocytic reactivity around the implant was assessed through immunofluorescent staining targeting glial fibrillary acidic protein (GFAP). The implants were removed, and 30 μm thick coronal sections of implanted brains were cut and incubated overnight with primary antibodies GFAP and NeuN, conjugated with secondary antibodies FITC-488nm and Cy3-532nm to label astrocytes and neurons, respectively, depicted in green and red within figures 7e and 7f. The soma of all cell types was stained with Hoechst-460 nm. The high density of neuronal cell bodies (NeuN and Hoechst) underlines the position of pyramidal neurons within the hippocampi (figure 7e-f). The slight increase of the GFAP staining reveals the position of the implant crossing the hippocampi. The increase in astrocytes around the implant, is significantly reduced compared to rigid implants, and results in part from the disruption of the tissue during the insertion of the implant. This effect could be reduced with the insertion speed in future studies. Although, the structures of the subcortical area remain extremely well-defined, such as the hippocampus observed within figure 7. The neuronal density remains mostly unaffected around the implant, with only localized network disruptions, indicating the absence of adverse reactions over the implantation time (58 days).

**Conclusion**

This paper describes the fabrication of a conformable, transparent and biocompatible intracortical microelectrode array and its characterization for electrical stimulation and recording, concomitant with hemodynamic brain measurements using fMRI. The implantation of a multichannel microelectrode array resulted in unobservable distortion and signal loss in fMRI images inside a 4.7-T MRI scanner. fMRI-induced electrical noise was removed using post-processing signal analysis allowing MRI data collection concomitant with spike recordings. These results show that we can combine the high spatiotemporal resolution and biocompatibility of flexible microdevices while maintaining a low magnetic cross-section for simultaneous whole-brain fMRI measurements. With these combined recordings, both large- and small-scale neural activity could be probed simultaneously allowing for examination of correlations between scales and brain functions in many situations. The electrical performances encompass planar and rigid devices and allow for single unit recordings during both acute and chronic recording sessions. Moreover, both the IRM and immunofluorescent staining confirm the absence of adverse reactions over time (58 days) and the high biocompatibility of the implant. Future investigations should provide dedicated connection strategy enabling to further measure tissue parameters such as anisotropic diffusion coefficient and susceptibility-weighted MRI sequences, and track any damage that can be attributed to the long-term MRI exposure in combination of intracortical probes.

**Materials and Methods**

***Device fabrication***. First, a bottom insulation layer of 10 μm Parylene-C is deposited on a clean Si wafer. After the first Parylene layer deposition, a photoresist was spin-coated and exposed with laser lithography to define metallic wires, contacts, and electrodes. Metals are then deposited in high vacuum ($10^{-7}$ bar) by evaporation of 5 nm of titanium and 45 nm of gold, lifted in acetone, and inspected under a standard microscope to check quality of the 5 µm wide contact lines. Then, the probes are again coated with 10 µm of Parylene for isolation during liquid operation. The devices were again cleaned, coated with photoresist (S1805), and patterned to outline the device shank. A 40 nm-thick aluminum layer was then evaporated over the devices as an etch mask, and the devices placed in RIE oxygen plasma for 120 minutes until all the Parylene is etched. The aluminum mask is removed using MF-319 developer, and the devices can be detached from the wafer after immersion in deionized water.

For graphene devices, high-quality CVD-grown graphene monolayers were transferred on the Parylene-C coated Si-wafer, using PMMA as carrier layer during the wet etching of coper foil in an ammonium persulfate solution. The graphene/PMMA stack was rinsed with 6 successive baths in DI water, then delicately fished with the silicon wafer and finally PMMA is removed in an overnight acetone bath. A gold hard mask was used during the reactive ion etching in oxygen plasma to define the graphene recording sites. Graphene devices are then connected to metallic leads obtained by the lift-off of 5 nm Ti, 25 nm Pd and 20 nm Au evaporated on a photoresist previously patterned with high-quality laser lithography. As reported for the metallic microelectrodes, a second layer of parylene-C is evaporated to isolate the contact lines for liquid operation and etched in RIE oxygen plasma to expose the graphene microelectrode to air and define the probe shank. During that final etching, the graphene layer is protected with an aluminium hard mask which is removed after etching the Parylene in resist developer (MF319, tetramethylammonium hydroxide and Polyalkylene glycol in DI water).

The device leads are designed to match the zero-insertion force (ZIF) connector (Würth Elektronik WR-FPC). In addition, some devices are also fastened directly to flat flexible cable (FFC) fixed with silver paste to ensure a reliable contact, and polyimide tape for mechanical strength. This direct connection to the FFC outperforms the baseline measured with the micromanipulator of the probe station (blue versus black curves figure S1b). The ZIF connection (red curve) has slightly higher impedance than the reference, but still performs equally or better than without connector.

***In-Vivo measurements***. Long Evans male rats (N = 3) were used for these experiments. They were kept in the same standard environmental conditions (12-h light/dark cycle) at a constant temperature of 22 °C, with food and water provided ad libitum. The experiments were carried out in accordance with the policy of Lyon 1 University, the Grenoble Institute des Neurosciences (GIN) and the French legislation. Experiments were performed in compliance with the European Community Council Directive of November 24, 1986 (86/609/EEC). The research was authorized by the Direction Départementale des Services Vétérinaires de l'Isère - Ministère de l'Agriculture et de la Pêche, France (Coizet, Véronique, PhD, permit number 381003). Every effort was made to minimize the number of animals used and their suffering during the experimental procedure. All procedures were reviewed and validated by the "Comité éthique du GIN n°004" agreed by the research ministry.

For surgery and electrophysiological recordings, rats (N = 3, weight: 365–450 g) were anaesthetized with an intra-peritoneal injection of ethyl carbamate (urethane, 1,25 g/kg, Sigma-Aldrich) and placed in a stereotaxic frame. The body temperature was maintained at 37°C with a thermostatic heating blanket throughout the recording sessions. A craniotomy was then performed to allow access to the structures of interest. Extracellular voltage excursions during acute monitoring were amplified, band-pass filtered (300 Hz–10 kHz), digitized at 10 kHz and recorded directly onto computer disc using a Micro 1401 data acquisition system (Cambridge Electronic Design [CED] Systems, Cambridge, UK) running CED data capture software (Spike 2). For muti-channels and chronic recording acquisition systems from multichannel system was used (sampling rate 40kHz).

***Extracellular multi-unit recordings*** were obtained on the left hemisphere. The electrode was lowered into the brain at the following coordinates (AP): - 4.68 mm caudal to Bregma; (ML): + 2.8mm lateral to midline and (DV): - 4.6 mm ventral to the brain surface according to the rat brain atlas[51]. To implant the flexible and conformable MEA, the probes were temporally reinforced with a rigid carrier to reliably insert them through the rodent's brain tissues. A fast-degradable glucose or PEG coating strategy was chosen, in which a PDMS mold is used to increase repeatability and prevent the probe from bending during the drying step. This method provided strong and consistent rigid coatings on the device, with a very high success rate and repeatability. The resulting sugar or PEG casing is 400 μm thick (+/- 100 μm) and approximately 100 μm wider than the device. Few seconds after the insertion, most of the coating dissolves in the brain fluid and residuals can be removed with a water-based solution. This restores all bio-suitable features of the probe, mainly its flexibility, conformability and biocompatibility that is expected to prevent inflammatory and immune-reactions that reduce device performance over time for chronic recordings. Note that the sugar coating has a small permanent effect on the electrical performance of the device, particularly at high frequencies (figure S7).

*Histology*. At the end of the experiment, animals were killed by an overdose of pentobarbital and perfused with 0.9 % saline followed by 4 % paraformaldehyde. Brains were removed and post fixed overnight in 4 % paraformaldehyde at 4 °C, before being transferred into sucrose for 36 h. Astrocytic reactivity and neuronal cell bodies were revealed with immuno-fluorescent staining against glial fibrillary acidic protein (GFAP) and NeuN respectively. Briefly, serial coronal (30 µm) sections were cut and incubated overnight with the anti-GFAP (1:1000, Dako) and anti-NeuN (1:500, AbCys) primary antibodies, rinsed in two PBS baths, incubated few hours with secondary-antibodies (1:200, FITC-488nm, and Cy3-532nm, Alexa fluor respectively) and lastly stained with Hoechst (1:2000, 460 nm). The brain slices were then observed with a fluorescence microscope (Olympus BX51).

*MRI imaging and in-situ electrical recordings* were performed at the IRMaGe MRI facility (Grenoble, France) with a 4.7 T Bruker Biospec system with Avance III electronics. A quadrature birdcage coil was used for RF transmission. A 4-channel receive array was used for signal reception to image the agar phantom, and an open single-channel rat head coil was used for rats. MRI gradient-echo (GE) images of electrodes and parts of the electrode systems embedded in gelled agarose (2 wt% agarose, 0.3 µmol/ml Dotarem) were acquired to assess their potential to induce susceptibility artefacts in the images. Imaging for electrode susceptibility was performed using a $T_1$-weighted GE fast low-angle shot (FLASH) sequence with (TR/TE/α = 50 ms/ 8 ms/20°) and a dual-echo GE sequence for $B_0$-mapping with TR/TE/α = 35 ms/1.2 and 4.1 ms/30°. Electrical recordings and stimulation with the MEA during MRI posed several challenges. The strong static magnetic field and switched magnetic field gradients potentially interferes with electronic equipment, preventing to position the pre-amplifier close to the microelectrode. Shielding reduces electrical interference, but it is ineffective against magnetic interference at low frequencies. Thus, the amplification stages and acquisition systems needed to be some distance away, facing the problem of the increasing capacitance of the cable between the electrode and the pre-amplifier (1-3 m) that acts as a voltage divider with the electrode impedance. To reduce signal loss, we measured current instead of voltage, which is insensitive to cable length.

RF heating was assessed in a gelled agar tube with the connected implant, at the position of the connector and at the tip of the MEA, using two high-precision MRI-compatible fiber-optic temperature probes (Photon Control Inc., Richmond, BC, USA) during a Turbo-RARE sequence exhibiting high specific absorption rate (TR 1.5 s, TE 39 ms, 11 slices of 512x512 voxels, 70x70 µm² in-plane voxel size, acquisition time 8 min 24 s). Measurements were performed both in presence of the implant and connector, as well as after removal of the implant, with identical positions for the temperature probes. A linear baseline acquired over a period of 4 minutes prior to the MRI sequence was removed from data. Temperature increase during the MRI sequence was modeled linearly. 20 s of data prior to and after the start of the sequence were excluded from the analysis to avoid any impact of a potential mismatch in the synchronization between temperature measurements and MRI acquisitions. The difference in slope between baseline and MRI sequence allows to measure the local SAR at the position of each probe.

During in vivo MRI, the ZIF connector used for the chronically implanted animals produced strong susceptibility artifacts, precluding good visualization of the implanted brain region in MRI. To avoid this artifact, a 3D radial ZTE MRI acquisition was performed to assess the visibility of the MEA. A volume of 256 x 256 x 256 mm³ was acquired with an isotropic voxel size of 157 μm (TR = 2.1 ms, flip angle 3°, acquisition time 7 min 14 s).

**References**

1. Hong, G. & Lieber, C. M. Novel electrode technologies for neural recordings. *Nat Rev Neurosci* **20**, 330–345 (2019).
2. Logothetis, N. K., Pauls, J., Augath, M., Trinath, T. & Oeltermann, A. Neurophysiological investigation of the basis of the fMRI signal. *Nature* **412**, 150–157 (2001).
3. Oeltermann, A., Augath, M. A. & Logothetis, N. K. Simultaneous recording of neuronal signals and functional NMR imaging. *Magnetic Resonance Imaging* **25**, 760–774 (2007).
4. Dunn, J. F. *et al.* Functional brain mapping at 9.4T using a new MRI-compatible electrode chronically implanted in rats. *Magnetic Resonance in Medicine* **61**, 222–228 (2009).
5. Tolias, A. S. *et al.* Mapping Cortical Activity Elicited with Electrical Microstimulation Using fMRI in the Macaque. *Neuron* **48**, 901–911 (2005).
6. Brázdil, M. *et al.* Reorganization of language-related neuronal networks in patients with left temporal lobe epilepsy – an fMRI study. *European Journal of Neurology* **12**, 268–275 (2005).
7. Kim, D. *et al.* An MRI-compatible, ultra-thin, flexible stimulator array for functional neuroimaging by direct stimulation of the rat brain. in *2014 36th Annual International Conference of the IEEE Engineering in Medicine and Biology Society* 6702–6705 (2014). doi:10.1109/EMBC.2014.6945166.
8. Lu, L. *et al.* Soft and MRI Compatible Neural Electrodes from Carbon Nanotube Fibers. *Nano Lett.* **19**, 1577–1586 (2019).
9. Zhao, S. *et al.* Graphene Encapsulated Copper Microwires as Highly MRI Compatible Neural Electrodes. *Nano Lett.* **16**, 7731–7738 (2016).
10. Paralikar, K. J. *et al.* Feasibility and safety of longitudinal magnetic resonance imaging in a rodent model with intracortical microwire implants. *J. Neural Eng.* **6**, 034001 (2009).
11. Lai, H.-Y., Younce, J. R., Albaugh, D. L., Kao, Y.-C. J. & Shih, Y.-Y. I. Functional MRI reveals frequency-dependent responses during deep brain stimulation at the subthalamic nucleus or internal globus pallidus. *NeuroImage* **84**, 11–18 (2014).
12. Tammer, R. *et al.* Compatibility of glass-guided recording microelectrodes in the brain stem of squirrel monkeys with high-resolution 3D MRI. *Journal of Neuroscience Methods* **153**, 221–229 (2006).

13. Santiesteban, F. M. M., Swanson, S. D., Noll, D. C. & Anderson, D. J. Magnetic resonance compatibility of multichannel silicon microelectrode systems for neural recording and stimulation: design criteria, tests, and recommendations. *IEEE Transactions on Biomedical Engineering* **53**, 547–558 (2006).
14. Jaime, S., Cavazos, J. E., Yang, Y. & Lu, H. Longitudinal observations using simultaneous fMRI, multiple channel electrophysiology recording, and chemical microiontophoresis in the rat brain. *Journal of Neuroscience Methods* **306**, 68–76 (2018).
15. Jorfi, M., Skousen, J. L., Weder, C. & Capadona, J. R. Progress Towards Biocompatible Intracortical Microelectrodes for Neural Interfacing Applications. *J Neural Eng* **12**, 011001 (2015).
16. Weltman, A., Yoo, J. & Meng, E. Flexible, Penetrating Brain Probes Enabled by Advances in Polymer Microfabrication. *Micromachines* **7**, 180 (2016).
17. Khodagholy, D. *et al.* In vivo recordings of brain activity using organic transistors. *Nat Commun* **4**, 1575 (2013).
18. Lee, W. *et al.* Transparent, conformable, active multielectrode array using organic electrochemical transistors. *Proc Natl Acad Sci U S A* **114**, 10554–10559 (2017).
19. Yang, G. *et al.* PtNPs/PEDOT:PSS-Modified Microelectrode Arrays for Detection of the Discharge of Head Direction Cells in the Retrosplenial Cortex of Rats under Dissociation between Visual and Vestibular Inputs. *Biosensors* **13**, 496 (2023).
20. Zhuang, Q. *et al.* Wafer-patterned, permeable, and stretchable liquid metal microelectrodes for implantable bioelectronics with chronic biocompatibility. *Science Advances* **9**, eadg8602 (2023).
21. Schiavone, G. *et al.* Soft, Implantable Bioelectronic Interfaces for Translational Research. *Advanced Materials* **32**, 1906512 (2020).
22. Fallegger, F. *et al.* MRI-Compatible and Conformal Electrocorticography Grids for Translational Research. *Advanced Science* **8**, 2003761 (2021).
23. Wang, X. *et al.* A Parylene Neural Probe Array for Multi-Region Deep Brain Recordings. *J. Microelectromech. Syst.* **29**, 499–513 (2020).
24. Fan, B., Rodriguez, A. V., Vercosa, D. G., Kemere, C. & Robinson, J. T. Sputtered porous Pt for wafer-scale manufacture of low-impedance flexible microelectrodes. *J Neural Eng* **17**, 036029 (2020).
25. Kuzum, D. *et al.* Transparent and flexible low noise graphene electrodes for simultaneous electrophysiology and neuroimaging. *Nat Commun* **5**, 1–10 (2014).
26. Masvidal-Codina, E. *et al.* High-resolution mapping of infraslow cortical brain activity enabled by graphene microtransistors. *Nature Mater* **18**, 280–288 (2019).
27. Bakhshaee Babaroud, N. *et al.* Multilayer CVD graphene electrodes using a transfer-free process for the next generation of optically transparent and MRI-compatible neural interfaces. *Microsyst Nanoeng* **8**, 1–14 (2022).
28. Park, D.-W. *et al.* Graphene-based carbon-layered electrode array technology for neural imaging and optogenetic applications. *Nat Commun* **5**, 1–11 (2014).
29. Baz Khan, K. R., Al-Othman, A., Al-Nashash, H. & Al-Sayah, M. Mxene/Polydimethylsiloxane (PDMS) Based Implantable and Flexible Bioelectrodes for Neural Sensing. in *2023 Advances in Science and Engineering Technology International Conferences (ASET)* 1–4 (2023). doi:10.1109/ASET56582.2023.10180745.
30. Driscoll, N. *et al.* Two-Dimensional Ti3C2 MXene for High-Resolution Neural Interfaces. *ACS Nano* **12**, 10419–10429 (2018).
31. Takeuchi, S., Ziegler, D., Yoshida, Y., Mabuchi, K. & Suzuki, T. Parylene flexible neural probes integrated with microfluidic channels. *Lab on a Chip* **5**, 519–523 (2005).
32. Chen, C.-H. *et al.* Three-dimensional flexible microprobe for recording the neural signal. *JM3.1* **9**, 031007 (2010).
33. Tian, L. *et al.* Large-area MRI-compatible epidermal electronic interfaces for prosthetic control and cognitive monitoring. *Nat Biomed Eng* **3**, 194–205 (2019).
34. Cui, H., Xie, X., Xu, S., Chan, L. L. H. & Hu, Y. Electrochemical characteristics of microelectrode designed for electrical stimulation. *BioMedical Engineering OnLine* **18**, 86 (2019).
35. Zhang, Y. *et al.* MRI magnetic compatible electrical neural interface: From materials to application. *Biosensors and Bioelectronics* **194**, 113592 (2021).

36. Boutet, A. *et al.* Improving Safety of MRI in Patients with Deep Brain Stimulation Devices. *Radiology* **296**, 250–262 (2020).
37. Breckenridge, L. J. *et al.* Advantages of using microfabricated extracellular electrodes for in vitro neuronal recording. *Journal of Neuroscience Research* **42**, 266–276 (1995).
38. Nair, R. R. *et al.* Fine Structure Constant Defines Visual Transparency of Graphene. *Science* **320**, 1308–1308 (2008).
39. Veliev, F., Briançon-Marjollet, A., Bouchiat, V. & Delacour, C. Impact of crystalline quality on neuronal affinity of pristine graphene. *Biomaterials* **86**, 33–41 (2016).
40. Du, X. *et al.* Graphene microelectrode arrays for neural activity detection. *J Biol Phys* **41**, 339–347 (2015).
41. Kireev, D. *et al.* Versatile Flexible Graphene Multielectrode Arrays. *Biosensors* **7**, 1 (2017).
42. Koerbitzer, B. *et al.* Graphene electrodes for stimulation of neuronal cells. *2D Mater.* **3**, 024004 (2016).
43. Randles, J. E. B. Kinetics of rapid electrode reactions. *Discuss. Faraday Soc.* **1**, 11 (1947).
44. Jorcin, J.-B., Orazem, M. E., Pébère, N. & Tribollet, B. CPE analysis by local electrochemical impedance spectroscopy. *Electrochimica Acta* **51**, 1473–1479 (2006).
45. Rammelt, U. & Reinhard, G. On the applicability of a constant phase element (CPE) to the estimation of roughness of solid metal electrodes. *Electrochimica Acta* **35**, 1045–1049 (1990).
46. García-Larrea, L. *et al.* Electrical stimulation of motor cortex for pain control: a combined PET-scan and electrophysiological study. *PAIN* **83**, 259–273 (1999).
47. *ASTM F2182-19e Test Method for Measurement of Radio Frequency Induced Heating On or Near Passive Implants During Magnetic Resonance Imaging. ASTM International.* doi:10.1520/F2182- 19E01.
48. Kiyatkin, E. A. Brain temperature and its role in physiology and pathophysiology: Lessons from 20 years of thermorecording. *Temperature* **6**, 271–333 (2019).
49. Heim, M. *et al.* Combined macro-/mesoporous microelectrode arrays for low-noise extracellular recording of neural networks. *Journal of Neurophysiology* **108**, 1793–1803 (2012).
50. Pautrat, A. *et al.* Revealing a novel nociceptive network that links the subthalamic nucleus to pain processing. *eLife* **7**, e36607 (2018).
51. Paxinos, G. & Watson, C. *The Rat Brain in Stereotaxic Coordinates: Hard Cover Edition.* (Elsevier, 2006).

**Acknowledgments**
Authors acknowledge grants from the French National Agency of scientific Research under the projects ANR-18-CE42-0003 NANOMESH. Part of this work was performed on the IRMaGe platform, member of France Life Imaging network (grant ANR-11-INBS-0006).

**Funding**
French National Agency of scientific Research under the projects ANR-18-CE42-0003 NANOMESH

**Data and materials availability**
All data are available in the main text or the supplementary materials.

**FIGURES**

**FIGURE 1. Wafer-scale fabrication of the flexible and transparent neural implant.** (a) Fabrication process steps of the microelectrode array. From left to right: 20-μm thick Parylene-C coating on the substrate, Au (100nm) deposition and pattern on the microelectrodes and contact leads, second deposition of 10-μm thick conforming Parylene, hard mask (Al-50 nm) deposition and pattern to protect non-etched area, oxygen plasma etching of the exposed Parylene-C area, and peeling of the flexible implant. (b) Optical

micrographs (from left to right) showing the flexible gold and graphene devices after and before release from the substrate. (c) Images of the manufactured probes assembled with the recording connecters.

**FIGURE 2. Electrochemical impedance spectroscopy.** (a) Nyquist (blue) and bode (green) plots (circle) and fits (red line) of the electrochemical impedance of 30 μm wide microelectrodes against a 4 mm² Pt reference electrode, measured in PBS, at 500 mV peak-to-peak. (b) Equivalent electric circuit models and fitting parameters used for each impedance and phase spectra. The model includes a resistive element $R_0$ in series with two Randles circuits. The Randles circuits are composed of a constant phase element $C_{PE}$ in parallel with a charge transfer resistance $R_{CT}$ which is in series with a linear diffusion element $Z_W$.

**FIGURE 3. Dual electrical and MRI recordings.** MR images shown were collected using a 3D Flash MRI sequence, isotropic resolution of 0.2 mm. The entire volume encompasses 128x128x80 voxels (26x26x16 mm³). The images show three orthogonal views of the volume. (a) The 10-channels implant is embedded between two successively poured layers of agarose gel (1 %) without connector. The plane of the implant is orthogonal to the axis of the magnetic field (x-y plane). In the two upper images, the interface between the two layers of agarose gel is faintly visible. The bottom left image is a section at the level of that interface. While the interface is visible, likely due to the presence of free water, the implant is not visible at all. Scale bar: 5 mm. (b) The 10-channels implant with the final FFC connector (inserted along the axis of the magnetic field, with the implant parallel to the x-z plane). The ribbon cable and the tape covering the connection are visible in the image, but not the implant located beyond the connection. (c-d) Microelectrodes current timeline during a 3D-Flash and a 3D-field map (d) sequences, while applying voltage impulses (10ms long) with one of the 10 channels through the agarose gel. The bottom traces (c) shows the MEA response to the stimuli without (grey) and during (black) the MRI acquisition. Spike-like events being detected with similar shape, amplitude of the recorded spikes and noise baseline. (e-f) RF heating of the Flex-MEA and connector during MRI. Fiber-optic thermometry during a high-SAR MRI sequence within an agar tube in absence (e) and presence (f) of the flex-MEA with connector, at the location of the flex-MEA tip (blue curves) and at the connector (orange curves). The duration of the MRI sequence was 504 s (grey box). A linear baseline was fitted to a period of 4 min prior to the MRI sequence and the SAR during MRI was estimated based on a linear fit to the temperature increase.

**FIGURE 4. Benchmark study during acute intracortical recording.** The voltage-time traces of the gold microelectrode (a), tungsten wire (b) and glass pipette (c) are compared during acute monitoring of subcortical neurons. All time traces show the spontaneous activity of individual neurons characterized with single unit detections. Zoomed views on selected spikes illustrate the typical shape of the recorded spike train. The spikes detected with the microelectrode (panel a) show similar shape with varying amplitude ranging from 100 to 400 μV, which is as expected for unitary action potentials. The inter-spike interval increases over time from 20 to 100 ms for the highest spike amplitude (bottom timestamps).

**FIGURE 5. Multi-channel recordings (day 11)** acquired with the 10 microelectrodes of the flexible probe at day 11 after the implantation surgery. The voltage-time traces show the spontaneous and evoked activity patterns, acquired during resting period and after an auditory or sensory stimulus (panels a, b and c respectively). For each recording sequences, the 10 microelectrodes are monitored and a zoom view of selected recorded channels highlights the typical recorded signals. At each sound stimulus, the microelectrode recorded an oscillatory variation of the local field potential which is as expected for the auditory system, while sensory stimuli induce single and train of unitary spikes. Baseline noise level is 20 µV. Output signals were amplified, band-pass filtered (300 Hz – 10 kHz) and digitized at 10 kHz.

**FIGURE 6. Multi-channel recordings (day 19)** acquired with the 10 microelectrodes of the flexible probe at day 19 after the implantation surgery. Same experiment is conducted than shown in figure 4. The voltage-time traces show the spontaneous and evoked activity patterns, acquired during resting period and after an auditory or sensory stimulus (panels a, b and c respectively). For each recording sequences, the 10 microelectrodes are monitored and a zoom view of selected recorded channels highlights the typical recorded signals.

**FIGURE 7. MRI and tissue footprints of the Flex-MEA (day 58), seven weeks after their implantation.** (a-d) Images acquired using a 3D ZTE MRI sequence with isotropic spatial resolution of 157 µm. The red crosses indicate the expected position of the implant on the three images. (d) Bottom view of the cap, showing the dental cement and the position of the screws maintaining the cap, at the same scale as the transverse MRI section in (a). (e-f) Immunofluorescent micrographs of coronal sections (30 µm) processed after the recordings and MRI acquisition sessions and removal of the implanted probe. Coronal sections have been stained with anti-GFAP (FITC, 488nm, green), anti-NeuN (Cy3, 532nm, red) to stain the population of astrocytes (GFAP) around the implant and the population of neurons (NeuN) respectively. All cell bodies have been stained with Hoechst (460 nm, blue). The high density of neuron underlines clearly the position of pyramidal neurons in the hippocampus, while the expected position of the implant is revealed by a slight proliferation of astrocytes. Scale bar 400 µm.

**Supplementary Materials**

Supplementary figures (S1 to S8) and table S1 are available in the supplementary materials.

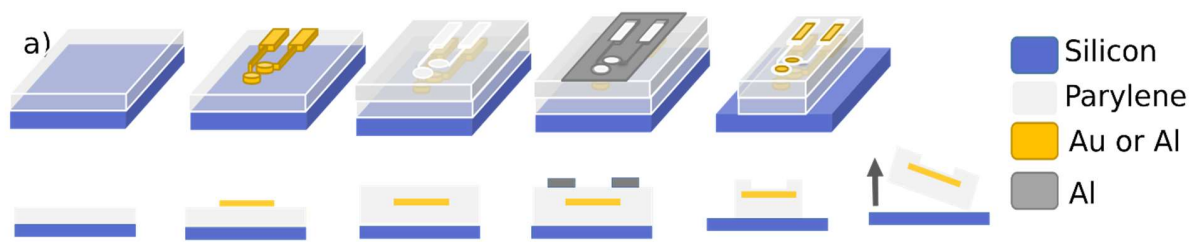
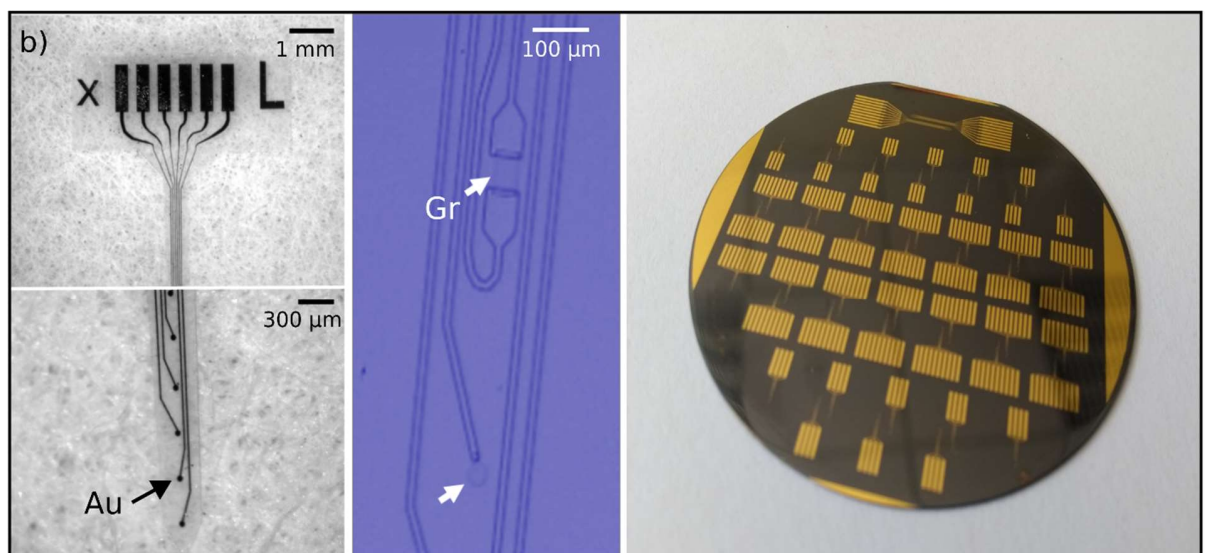
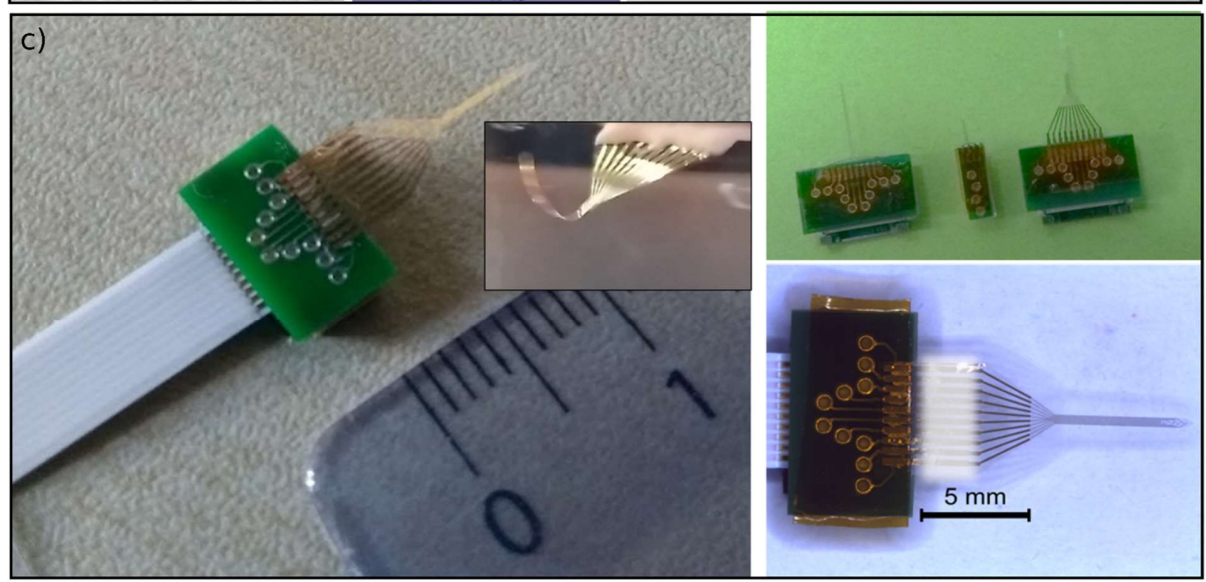

Figure 1.

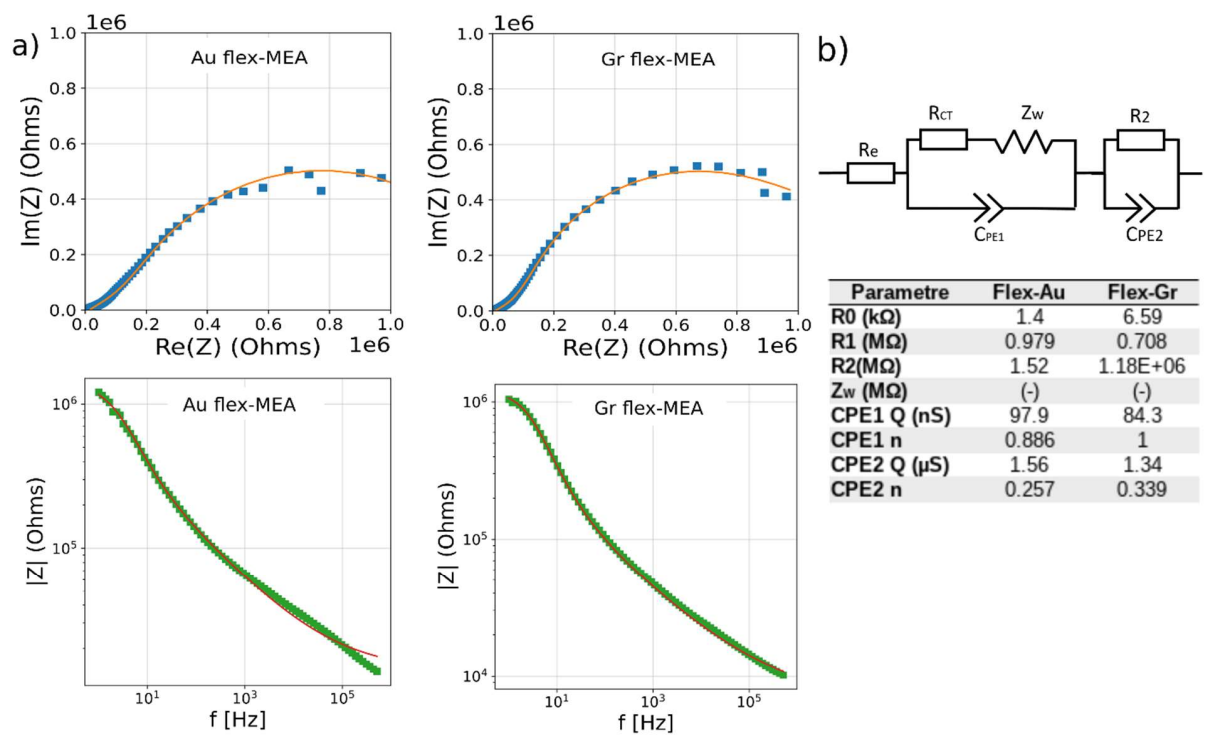

Figure 2.

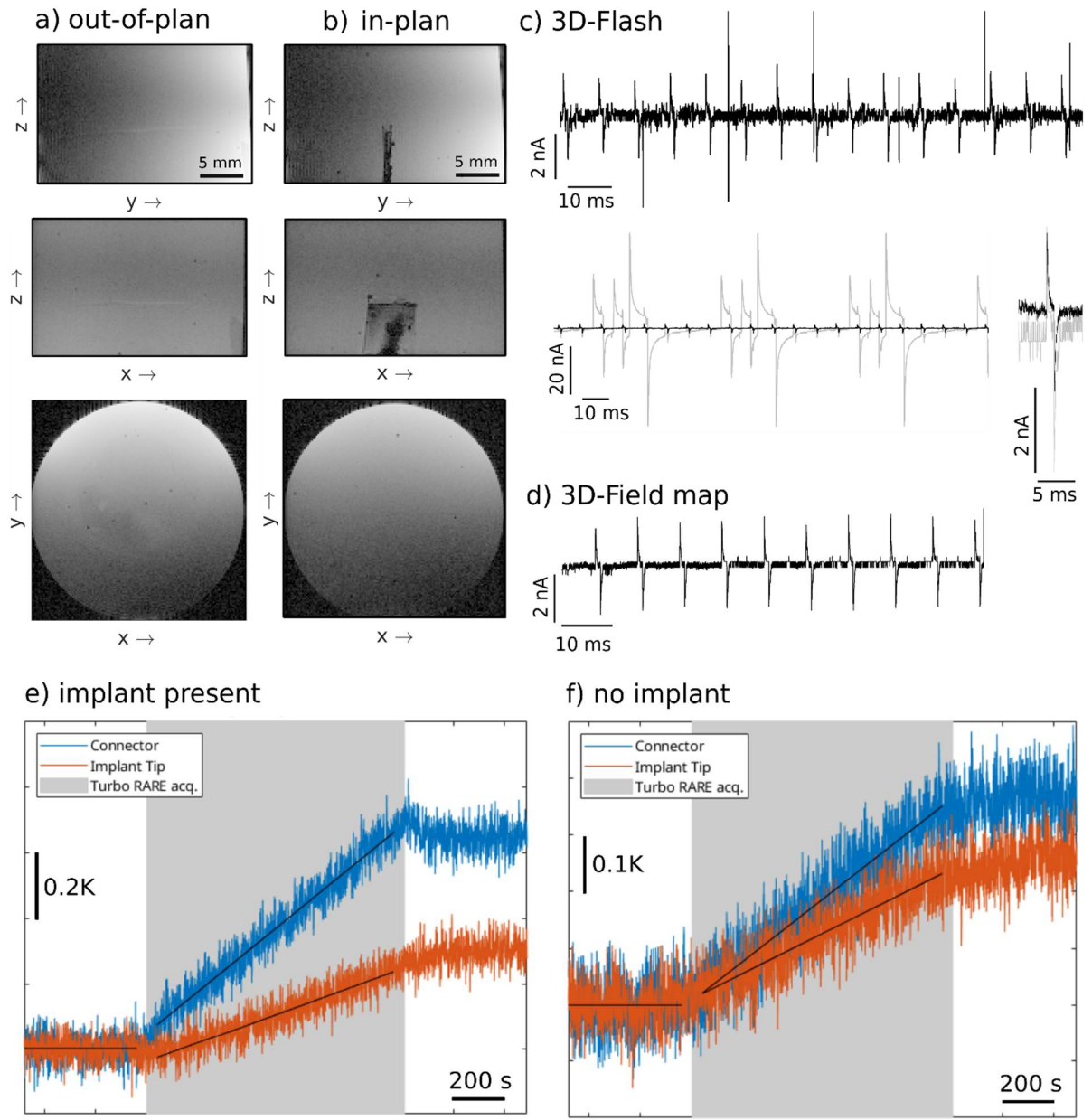

Figure 3.

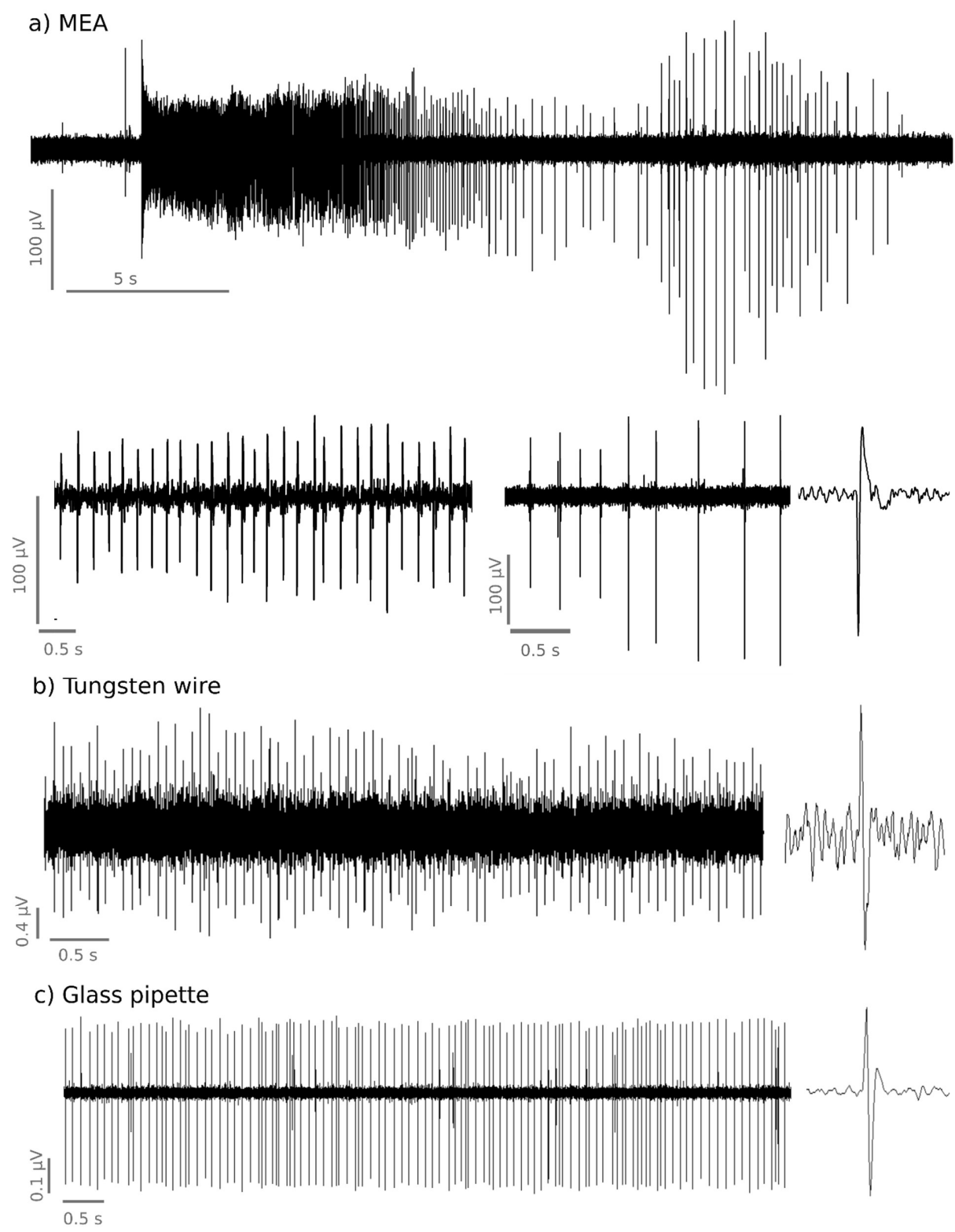

Figure 4.

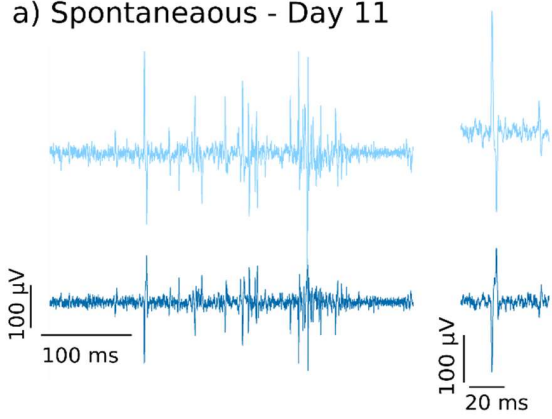
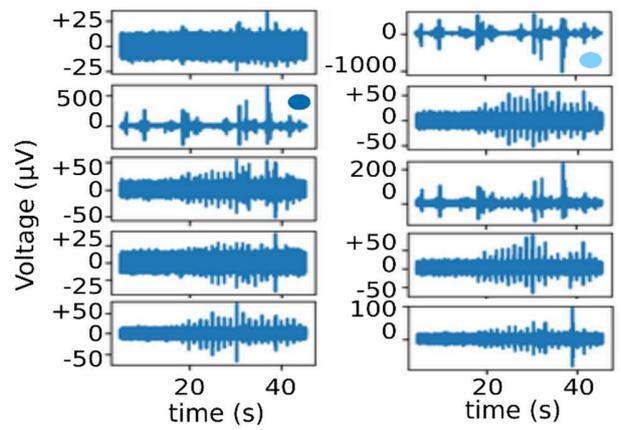

a) Spontaneous - Day 11

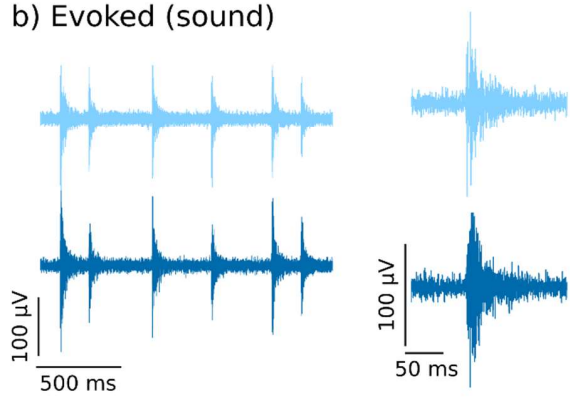
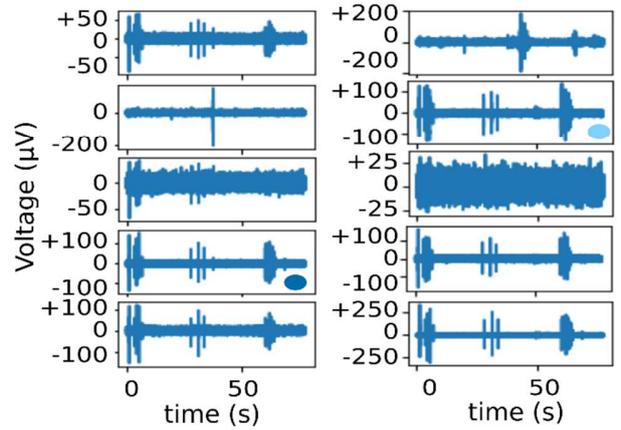

b) Evoked (sound)

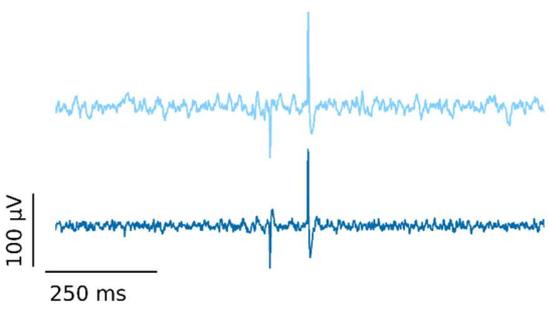
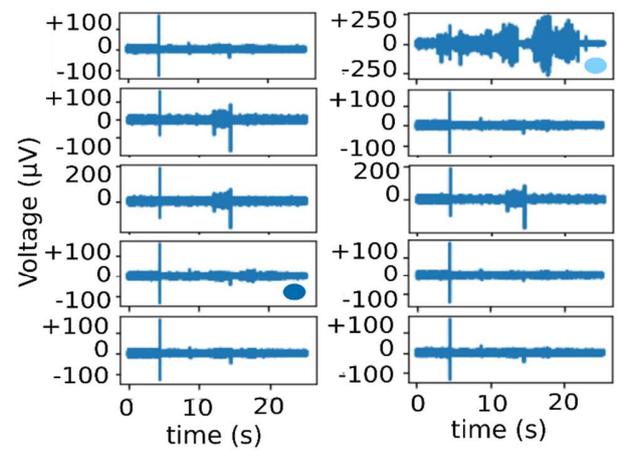

c) Evoked (sensory)

Figure 5.

a) Spontaneaous - Day 19

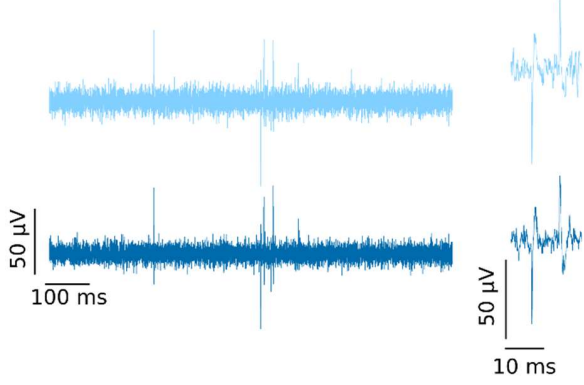
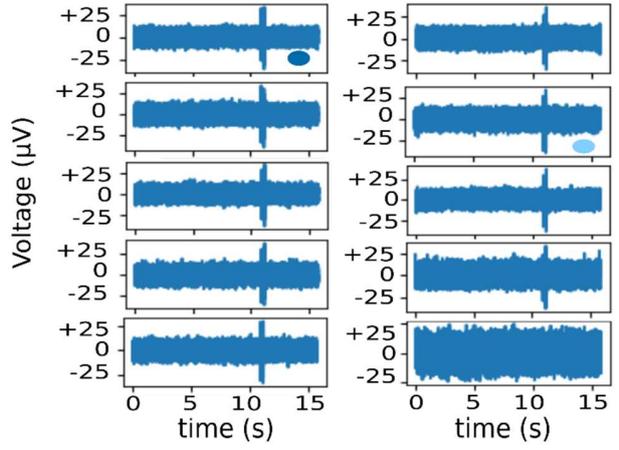

b) Evoked (sound)

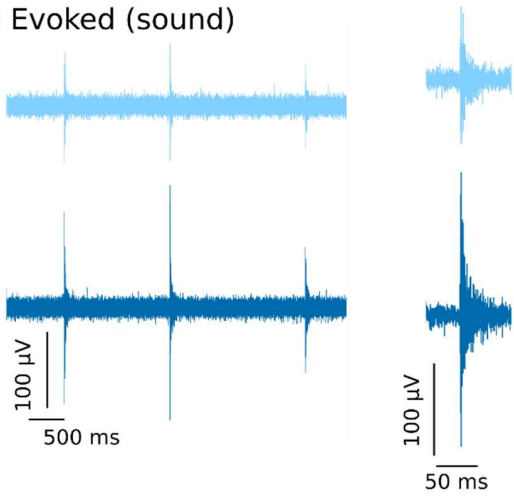
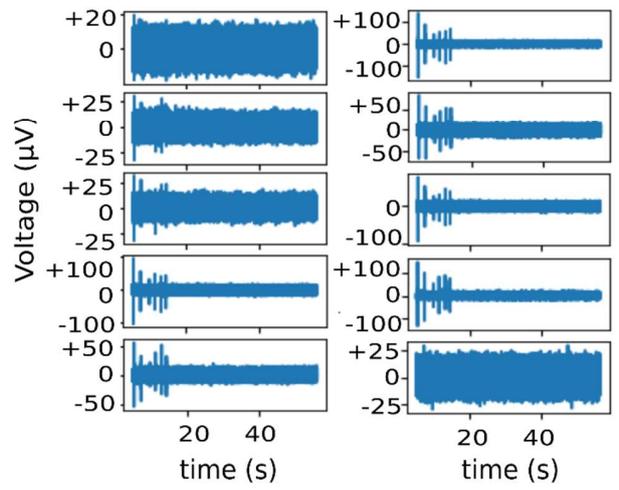

c) Evoked (sensory)

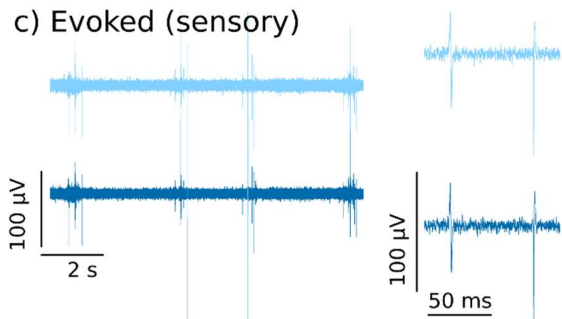
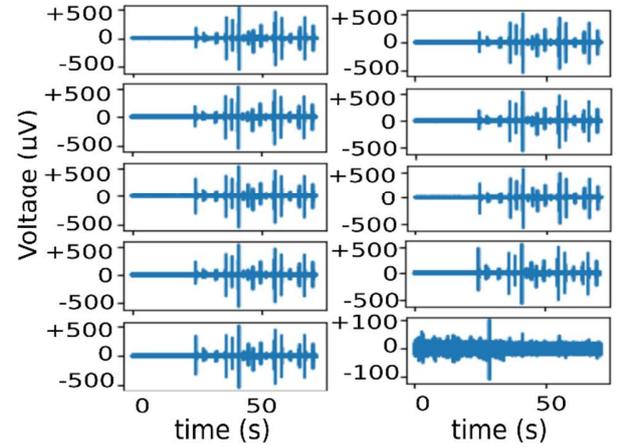

Figure 6.

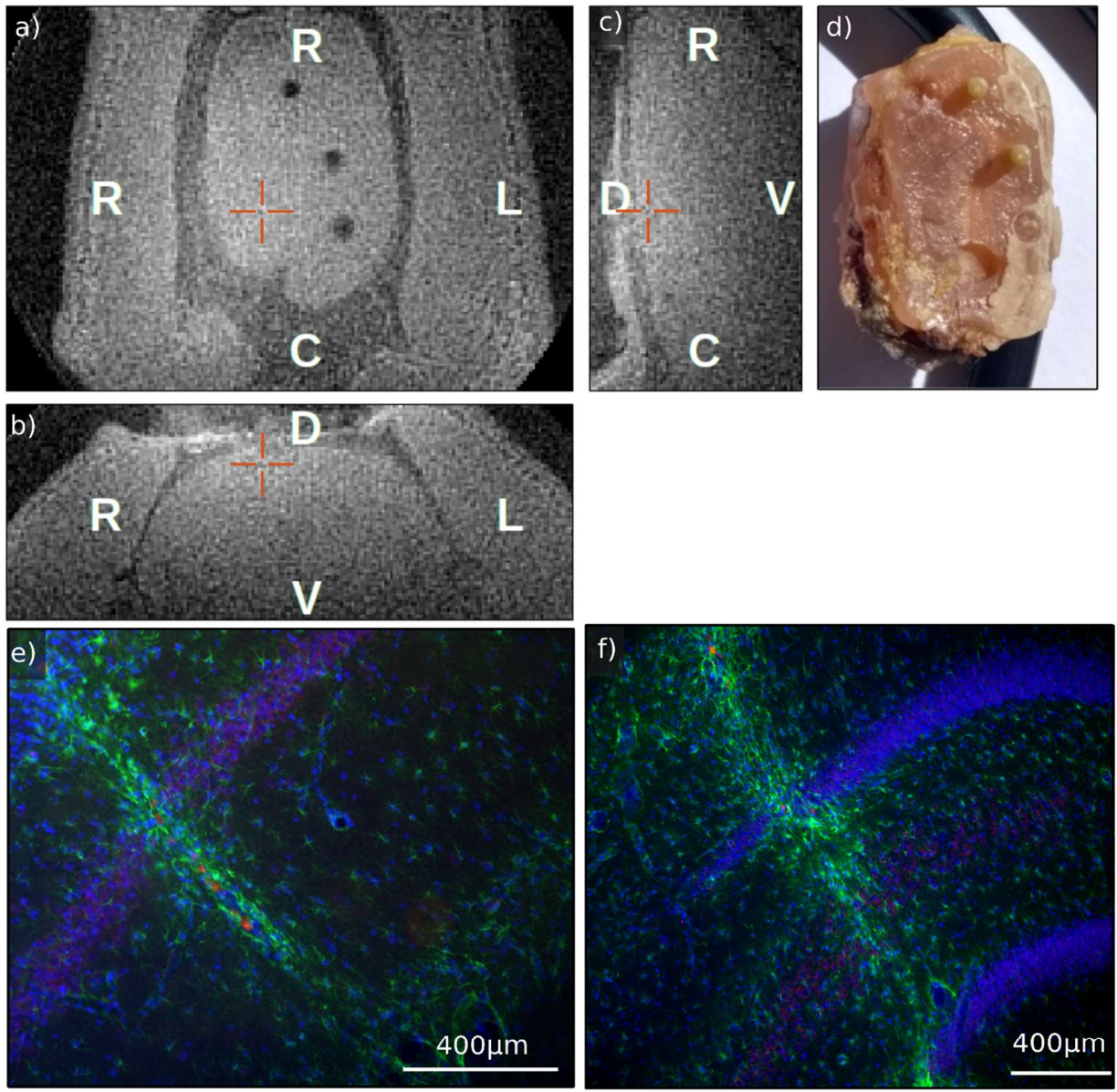

Figure 7.